\documentclass[11pt]{article}
\usepackage[margin=1in]{geometry}
\usepackage{graphicx}
\usepackage{subfigure}
\usepackage{makecell}
\usepackage{amsmath}
\usepackage{amssymb}
\usepackage{comment}
\usepackage{longtable}
\usepackage[hyperfootnotes=true]{hyperref}
\usepackage{float}
\hypersetup{
 colorlinks=true,
 citecolor=blue,
 linkcolor=blue,
 urlcolor=blue}

\def\beq{\begin{equation}}
\def\eeq#1{\label{#1}\end{equation}}
\def\eeqn{\end{equation}}

\def\beqa{\begin{eqnarray}}
\def\eeqa#1{\label{#1}\end{eqnarray}}
\def\eeqan{\end{eqnarray}}

\let\bar=\overbar

\def\Dslash{\not{\hbox{\kern-4pt $D$}}}
\def\dslash{\not{\hbox{\kern-2pt $\del$}}}

\def\msb{{\bar{\ssstyle M \kern -1pt S}}}

\def\Title#1{\begin{center} {\Large {\bf #1} } \end{center}}
\def\Author#1{\begin{center} {\normalsize {\sc #1} } \end{center}}
\def\Institution#1{\begin{center} {\normalsize {\it #1} } \end{center}}
\def\Abstract#1{\noindent {\normalsize {\bf Abstract:} {\normalfont #1}}}
\def\Conference{\vspace{4mm}\begin{raggedright} {\normalsize {\it Talk presented at the 2019 Meeting of the Division of Particles and Fields of the American Physical Society (DPF2019), July 29--August 2, 2019, Northeastern University, Boston, C1907293.} } \end{raggedright}\vspace{4mm}}

\begin{document}

%
%

\Title{The Measurement of Position Resolution of RD53A Pixel Modules}

\Author{Gang Zhang\textsuperscript{1}, Benjamin Nachman\textsuperscript{2}, Shi-Chieh Hsu\textsuperscript{3} and Xin Chen\textsuperscript{1}}

\Institution{\textsuperscript{1}Tsinghua University, Beijing, China \\ \textsuperscript{2}Lawrence Berkeley National Laboratory, Berkeley, USA \\ \textsuperscript{3} University of Washington, Seattle, USA}

\Abstract{Position resolution is a key property of the innermost layer of the upgraded ATLAS and CMS pixel detectors for determining track reconstruction and flavor tagging performance.  The 11 GeV electron beam at the SLAC End Station A was used to measure the position resolution of RD53A modules with a $50\times50$ and a $25\times100\ \mu$m$^2$ pitch.  Tracks are reconstructed from hits on telescope planes using the EUTelescope package.  The position resolution is extracted by comparing the extrapolated track and the hit position on the RD53A modules, correcting for the tracking resolution.  10.9 and 6.8 $\mu$m resolution can be achieved for the 50 and 25 $\mu$m directions, respectively, with a 13 degree tilt.
}

\Conference

%
%

\section{Introduction}

The LHC will be upgraded to the High Luminosity LHC (HL-LHC) with the instantaneous luminosity
up to about $5 \times 10^{34}$ cm$^{-2}$ s$^{-1}$~\cite{HL-LHC}. To cope with such an intense collision environment, the current ATLAS and CMS tracking detectors will be replaced.  The RD53A module is intended to demonstrate in 
a large format IC the suitability of the chosen 65 nm CMOS technology for the the HL-LHC pixel detectors~\cite{RD53A}. It will form the basis
for the production designs for ATLAS and CMS. Position resolution is a key property of a pixel detector, which can be measured
by testbeam campaigns. 

\section{The testbeam measurement at SLAC}

\begin{figure}[!htb]
\centering
\includegraphics[width=\textwidth]{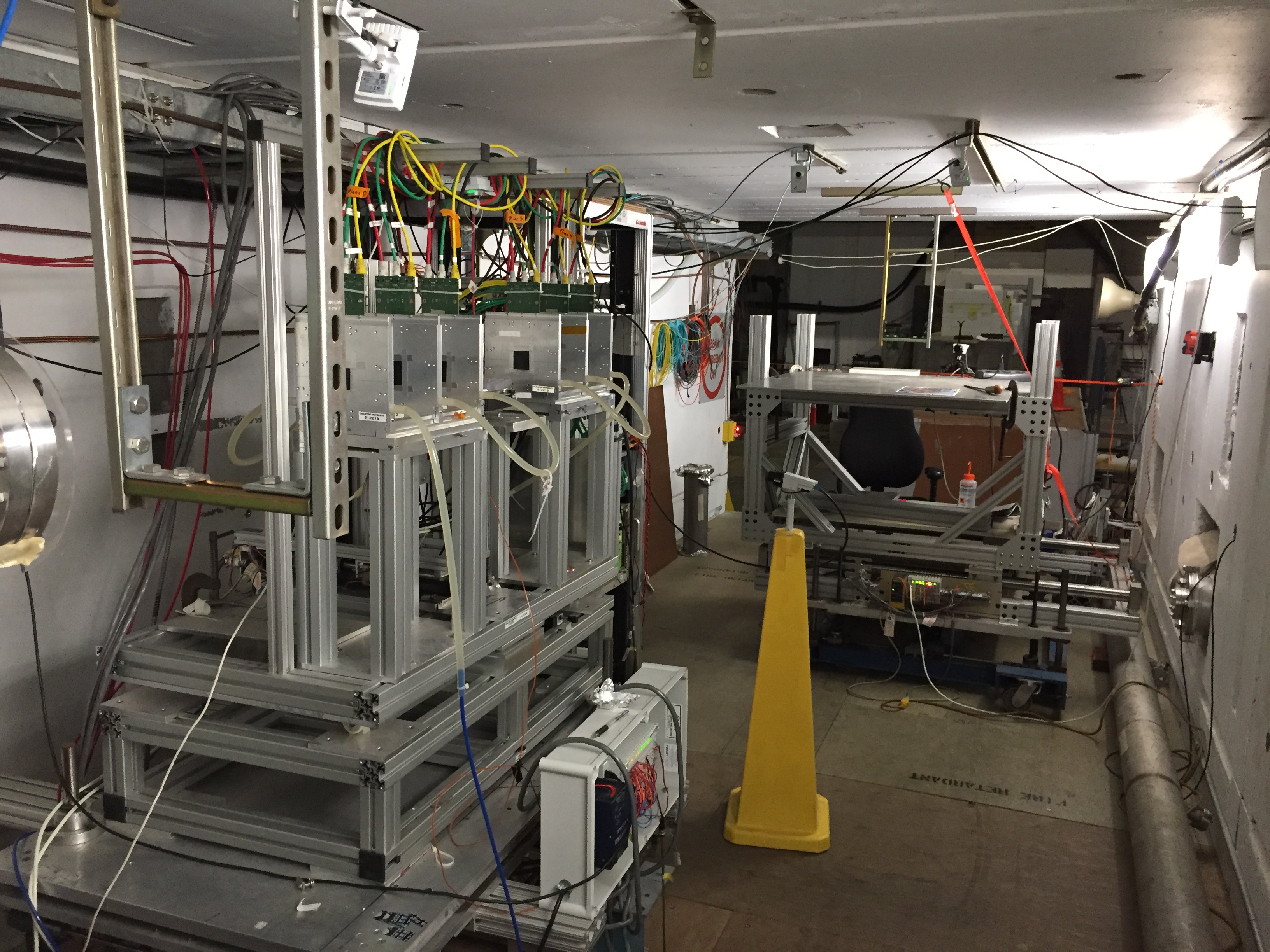}
\caption{Picture of SLAC testbeam set-up.}
\label{fig:telescopes}
\end{figure}

Testbeam measurements are important for developing new sensor and Front-End designs. In testbeam measurement, a controlled beam of charged particles are shot at the device under test (DUT), and the DUT's response to the incident particles can be studied.  A 5 Hz, 11 GeV electron beam at the SLAC End Station A was used in SLAC teatbeam. The data were collected by YARR~\cite{YARR} and synchronized by EUDAQ~\cite{EUDAQ} in November 2018. Figure~\ref{fig:telescopes} shows the layout of SLAC testbeam. The DESY EUDET-like telescope CALADIUM~\cite{intrinsic_reso} is used for reference sensors to reconstruct the track of the charged particles precisely. CALADIUM consists of six Mimosa26 sensors, three upstream and three downstream of the DUT. The Mimosa26 with 18.4 $\mu$m pitch monolithic active squared pixels is designed by IPHC in Strasbourg~\cite{Mimosa26}. There are 1152 rows and 576 columns in Mimosa26 with size $21.2\times10.6$ mm$^2$. Two kinds of DUTs are measured in SLAC testbeam. One is RD53A module with $50\times50\ \mu$m$^2$ pitch, 400 pixels in row and 192 pixels in column. The other one is RD53A module with $100\times25\ \mu$m$^2$ pitch, 200 pixels in row and 384 pixels in column. Their sizes are both $20.0\times9.6$ mm$^2$. For convenience to put the cables, the DUTs are rotated by $90^\circ$ in X-Y plane (Beam direction is Z axis and Y axis points upwards to form the right-hand coordinate system). The effect of a DUT tilt angle is also studied in this testbeam measurement where the DUTs are rotated by $13^\circ$ in the X-Z plane.

\section{Testbeam reconstruction and analysis}

The tool used to reconstruct the testbeam is EUTelescope v2.0.0~\cite{EUTelescope_1}\cite{EUTelescope_2}\cite{EUTelescope_3}, which is a collection of processors in Modular Analysis and Reconstruction for the Linear Collider (Marlin). The track reconstruction algorithm used is the General Broken Lines (GBL) fitter~\cite{GBL_1}\cite{GBL_2}. The procedures to process the testbeam data are similar to other testbeam studies and consist of five steps~\cite{testbeam}:
\begin{itemize}
	\item Data conversion: Transfer the raw sensor data into Linear Collider Input/Output (LCIO) format~\cite{LCIO}. In this step, the pixels with hit rate greater than given firing frequency are marked as noisy pixels and stored in a database.
	\item Clustering: Adjacent pixels are grouped together. Adjacent pixels mean they at least touch in the corner,  i.e. each pixel has eight neighbours. The clusters containing noisy pixels will be excluded.
	\item Hit derivation: The mean position of a cluster is computed based on binary readout, i.e. only considering the pixel is hit or not and ignoring charge weight. The pre-alignment is also conducted in this step. The pre-alignment processor assumes that tracks have a perpendicular incidence and thus hits in the global frame should not change in x- and y-direction, if multiple scattering is neglected. The differences of the hit position on the first telescope plane and all subsequent planes reflect the misalignments.
	\item Alignment: The pre-aligned hits are used to reconstruct the tracks using GBL fitter. Residuals from the track fit are passed to Millepede \uppercase\expandafter{\romannumeral2}~\cite{Millepede} to do the alignment. There are 3 iterations of the GBL-based alignment.
	\item Track fit: Only the aligned hits on the telescope are used to do the final fitting with the GBL algorithm.
\end{itemize}

\begin{figure}[!htb]      
\centering
\subfigure[]{
\label{fig:correlation:a} 
\includegraphics[width=0.45\textwidth]{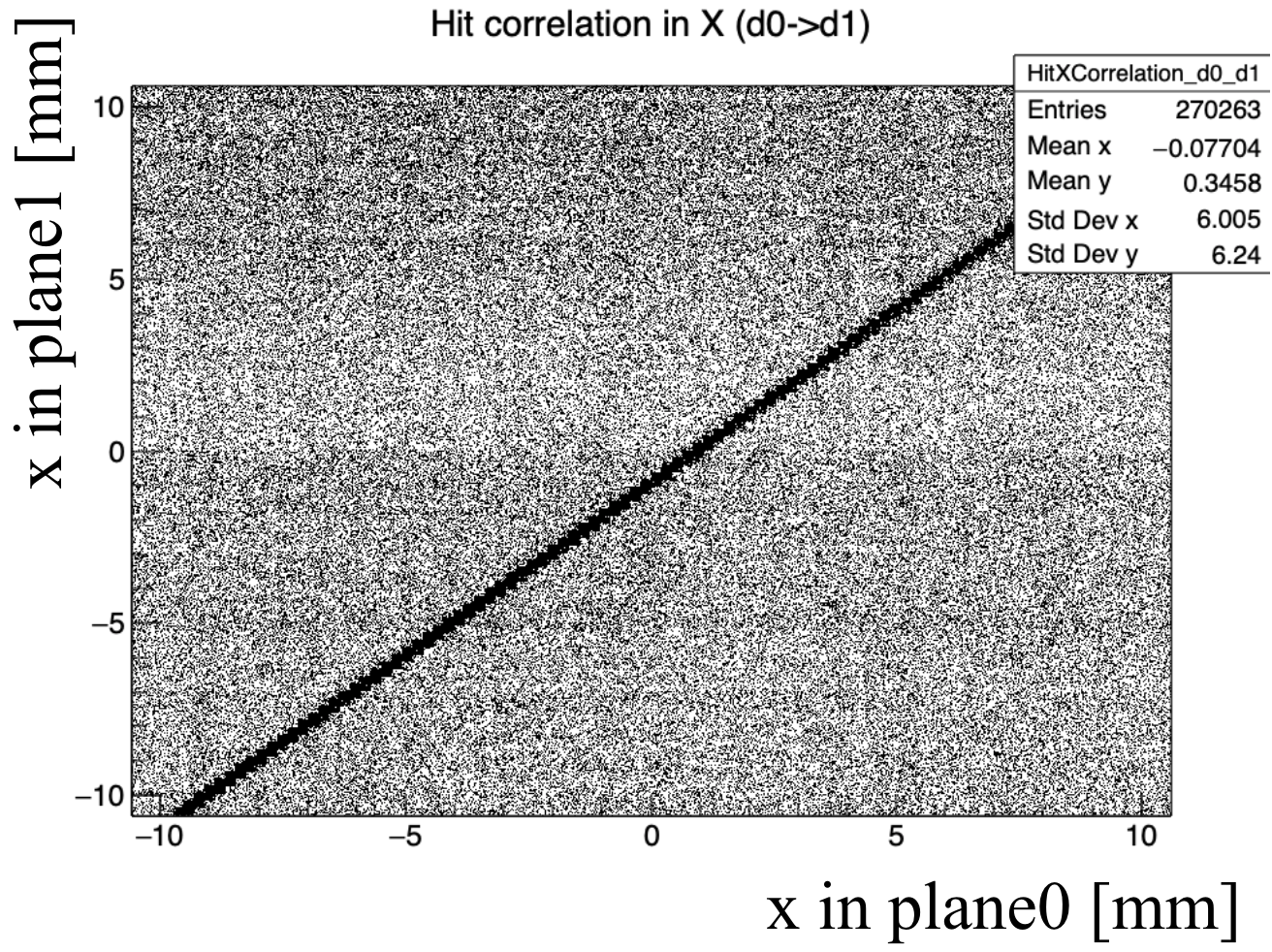}
}
\subfigure[]{
\label{fig:correlation:b} 
\includegraphics[width=0.45\textwidth]{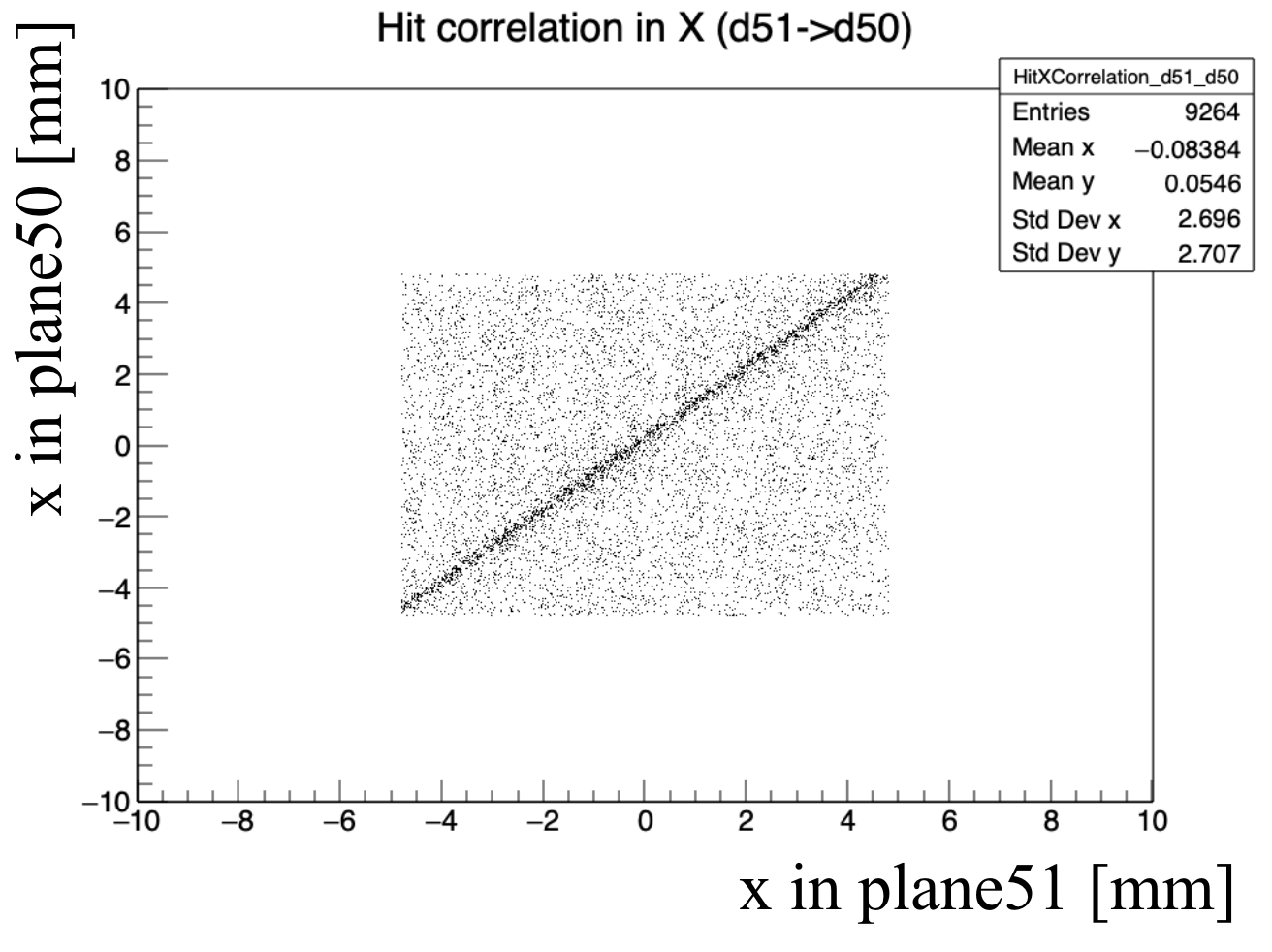}
}
\subfigure[]{
\label{fig:correlation:c} 
\includegraphics[width=0.45\textwidth]{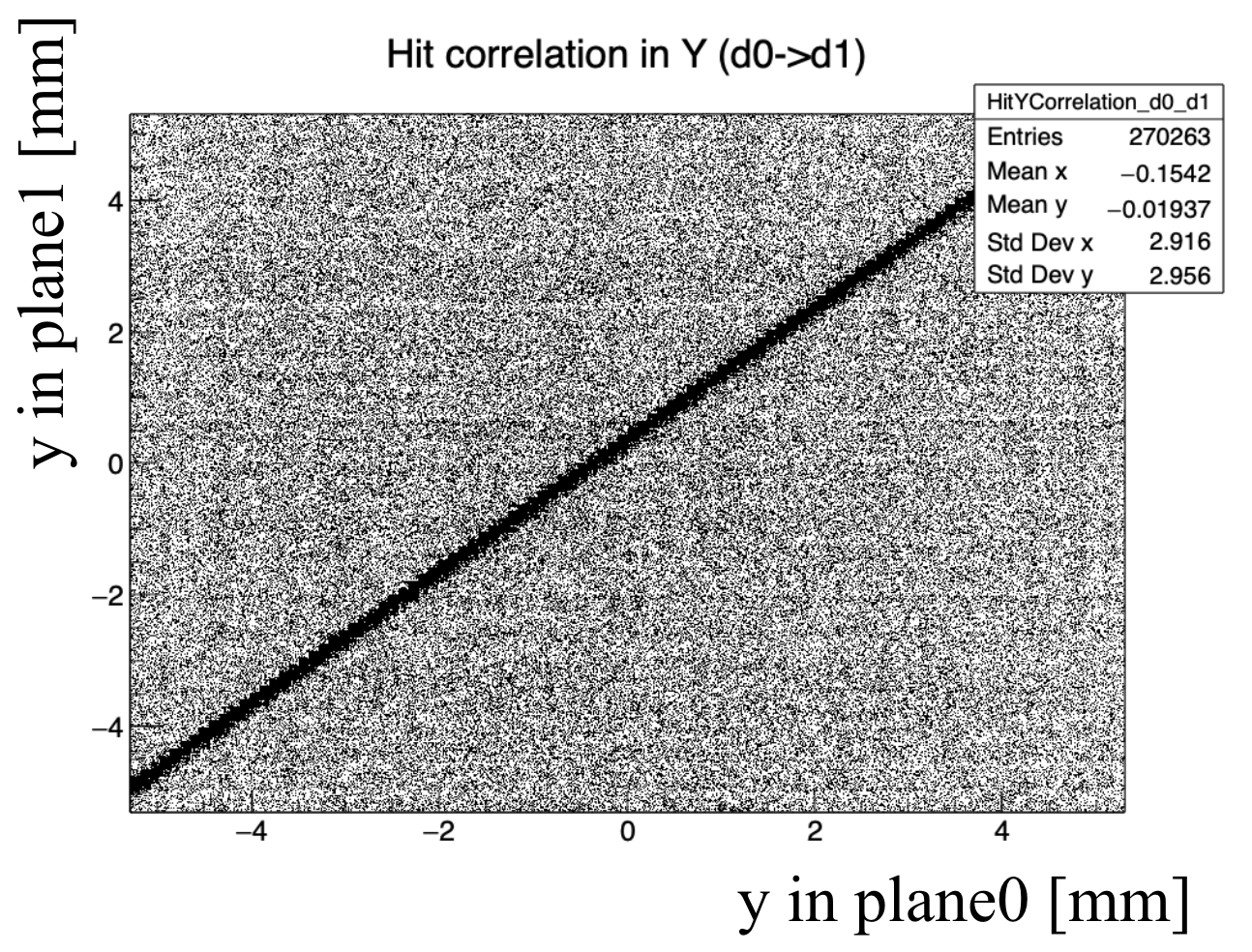}
}
\subfigure[]{
\label{fig:correlation:d} 
\includegraphics[width=0.45\textwidth]{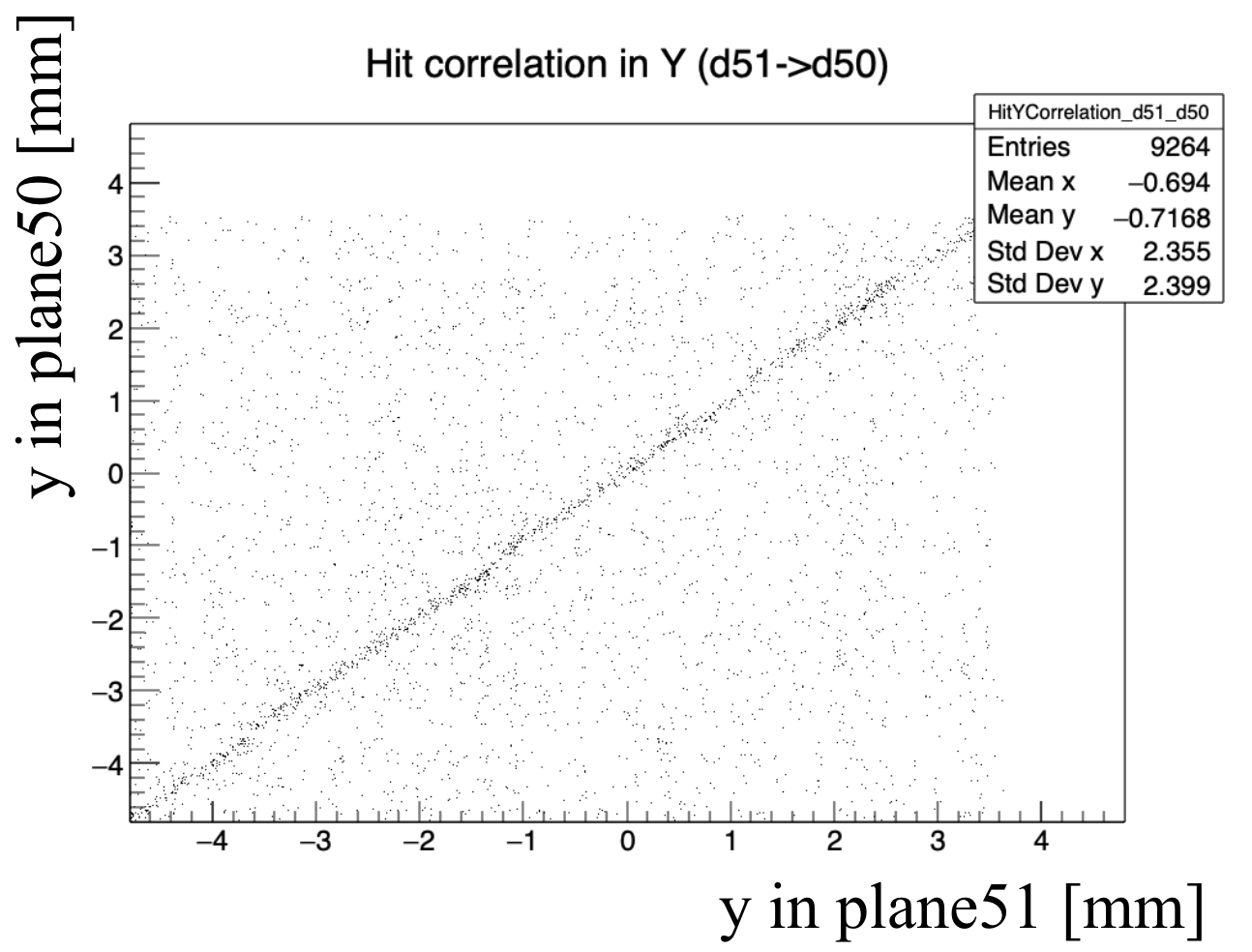}
}
\caption{Hits position correlations between two planes. (a) correlation in X direction between first and second Mimosa26. (b) correlation in X direction between two DUTs. (c) correlation in Y direction between first and second Mimosa26. (d) correlation in Y direction between two DUTs.}
\label{fig:correlation} 
\end{figure}

Figure~\ref{fig:correlation} shows the correlation of hit positions in the X and Y directions. The positive correlations demonstrate the geometry description and configuration in EUTelescope are correct. If there is misalignment between two planes, the correlation line will not pass through the origin.

\begin{figure}[!htb]      
\centering
\subfigure[]{
\label{fig:residual_Mimosa26:a} 
\includegraphics[width=0.45\textwidth]{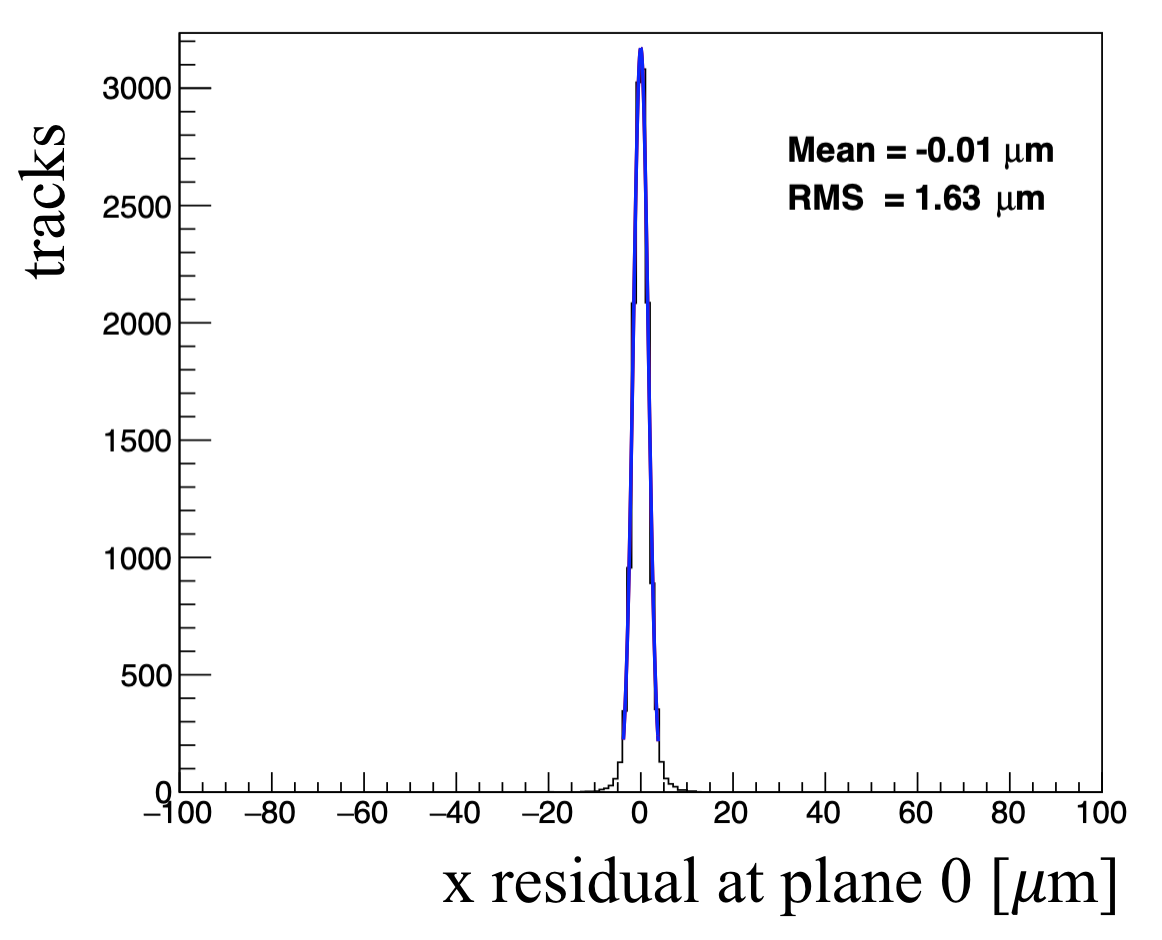}
}
\subfigure[]{
\label{fig:residual_Mimosa26:b} 
\includegraphics[width=0.45\textwidth]{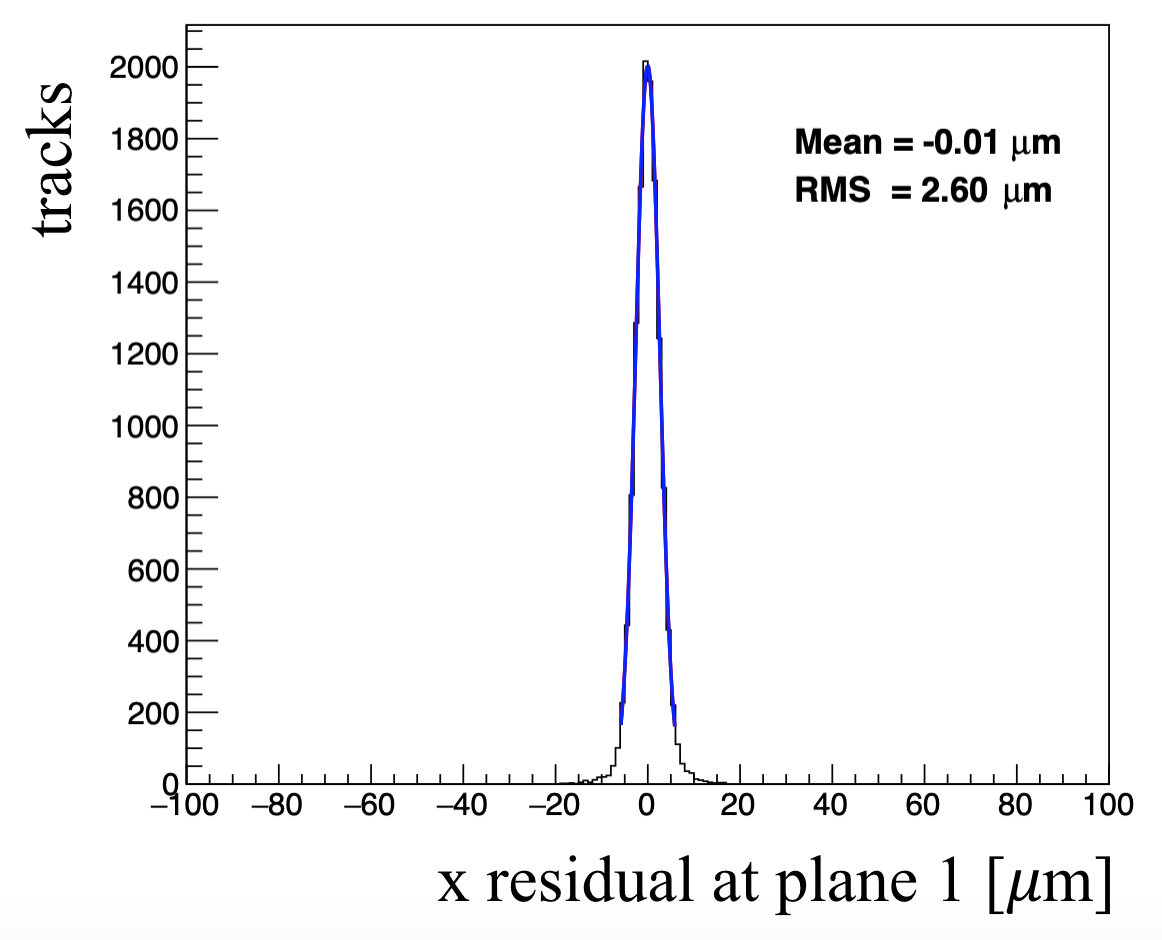}
}
\subfigure[]{
\label{fig:residual_Mimosa26:c} 
\includegraphics[width=0.45\textwidth]{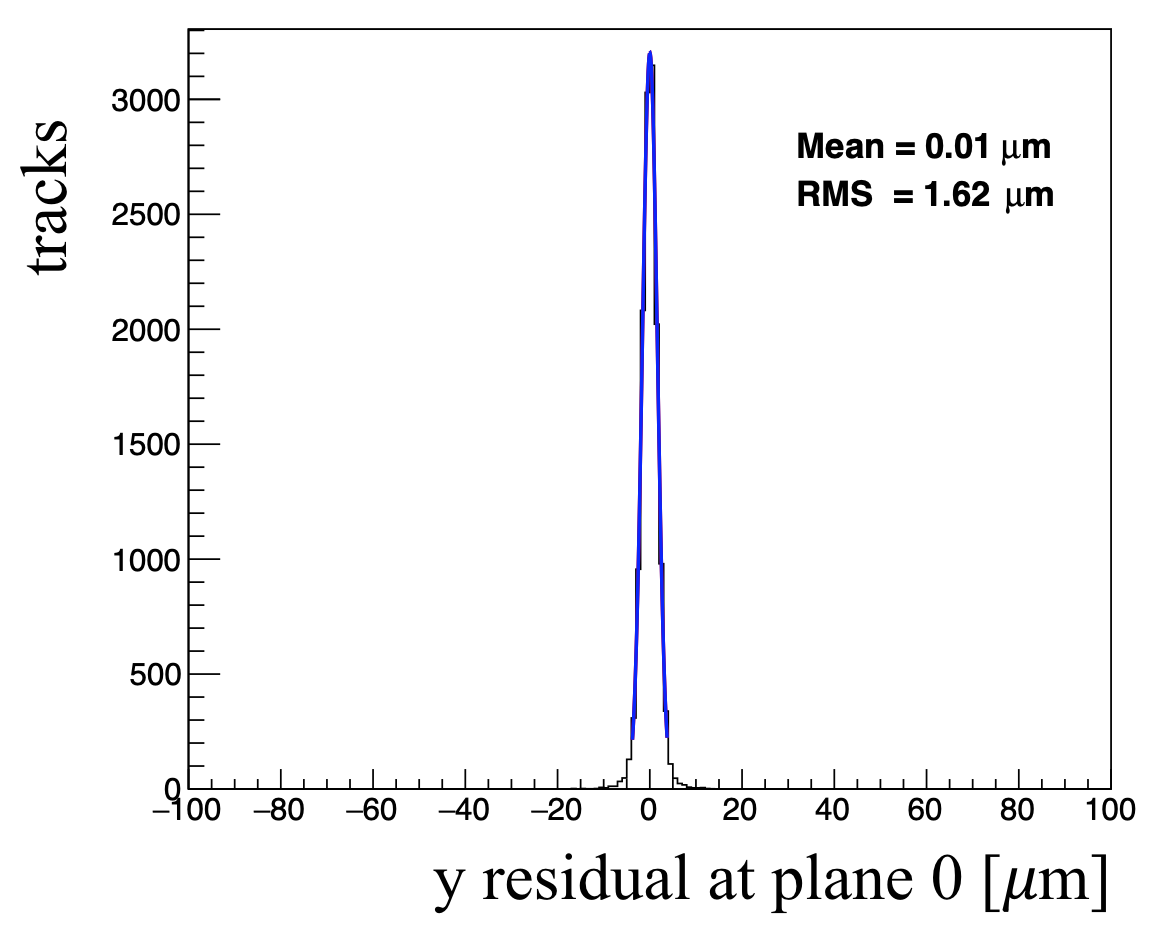}
}
\subfigure[]{
\label{fig:residual_Mimosa26:d} 
\includegraphics[width=0.45\textwidth]{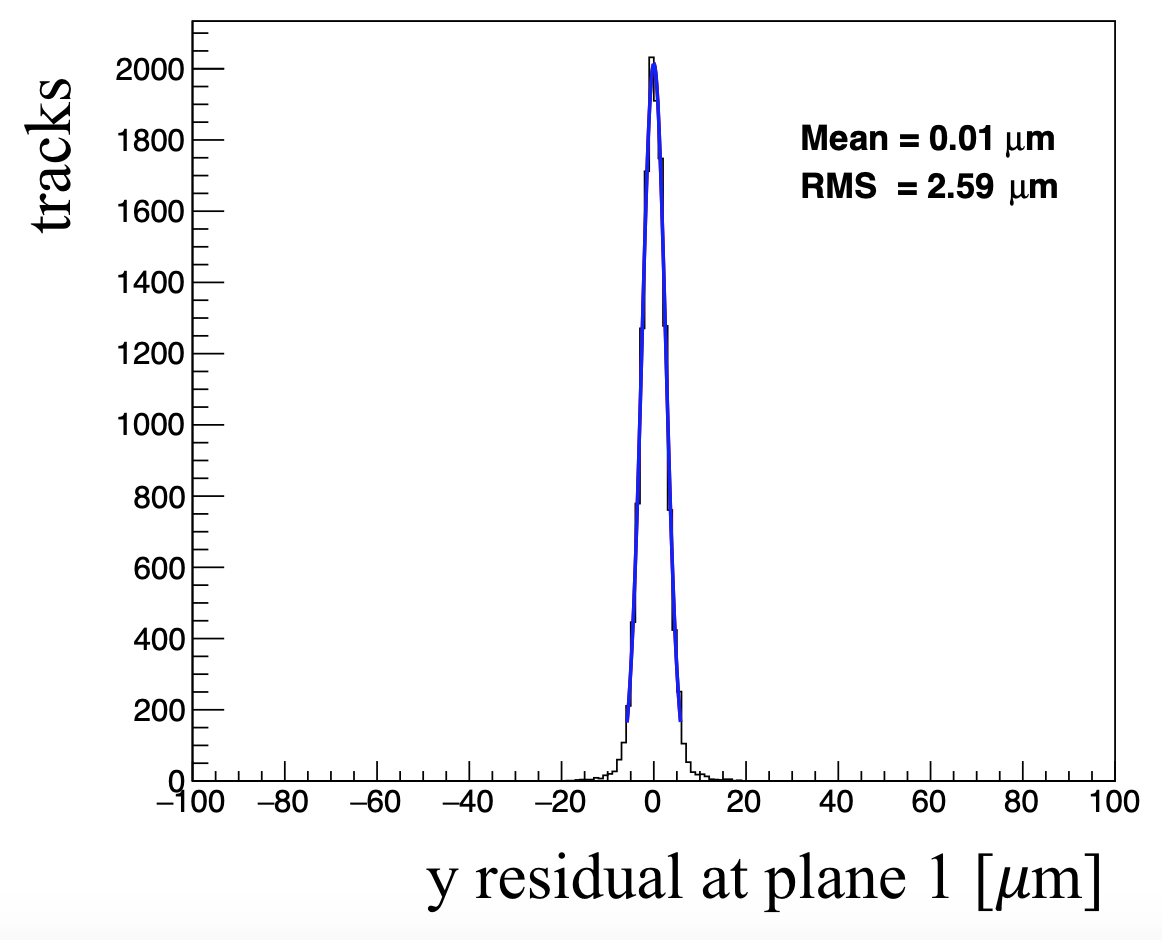}
}
\caption{Residual distribution of Mimosa26. The blue line is fitted gaussian function. (a) residual in X direction on first Mimosa26. (b) residual in X direction on second Mimosa26. (c) residual in Y direction on first Mimosa26. (d) residual in Y direction on second Mimosa26.}
\label{fig:residual_Mimosa26} 
\end{figure}

The residual, i.e. the difference between hit and reconstructed track position, on the first and second Mimosa26 are shown in Figure~\ref{fig:residual_Mimosa26}. A gaussian function is used to extract the position resolution of  Mimosa26. The residual distributions on other Mimosa26 are similar to these two planes. The small resolutions demonstrate the track reconstruction is good. Note that the residuals on Mimosa26 are biased since the hits on Mimosa26 are used in track fitting. Biased resolution is less than intrinsic resolution(Details in Appendix~\ref{apx:proof}):
\begin{equation}
\sigma^{2}_{biased}=\sigma^{2}_{intrinsic}-\sigma^{2}_{track}.
\end{equation}

\begin{figure}[!htb]      
\centering
\subfigure[]{
\label{fig:residual_RD53A_100_25:a} 
\includegraphics[width=0.45\textwidth]{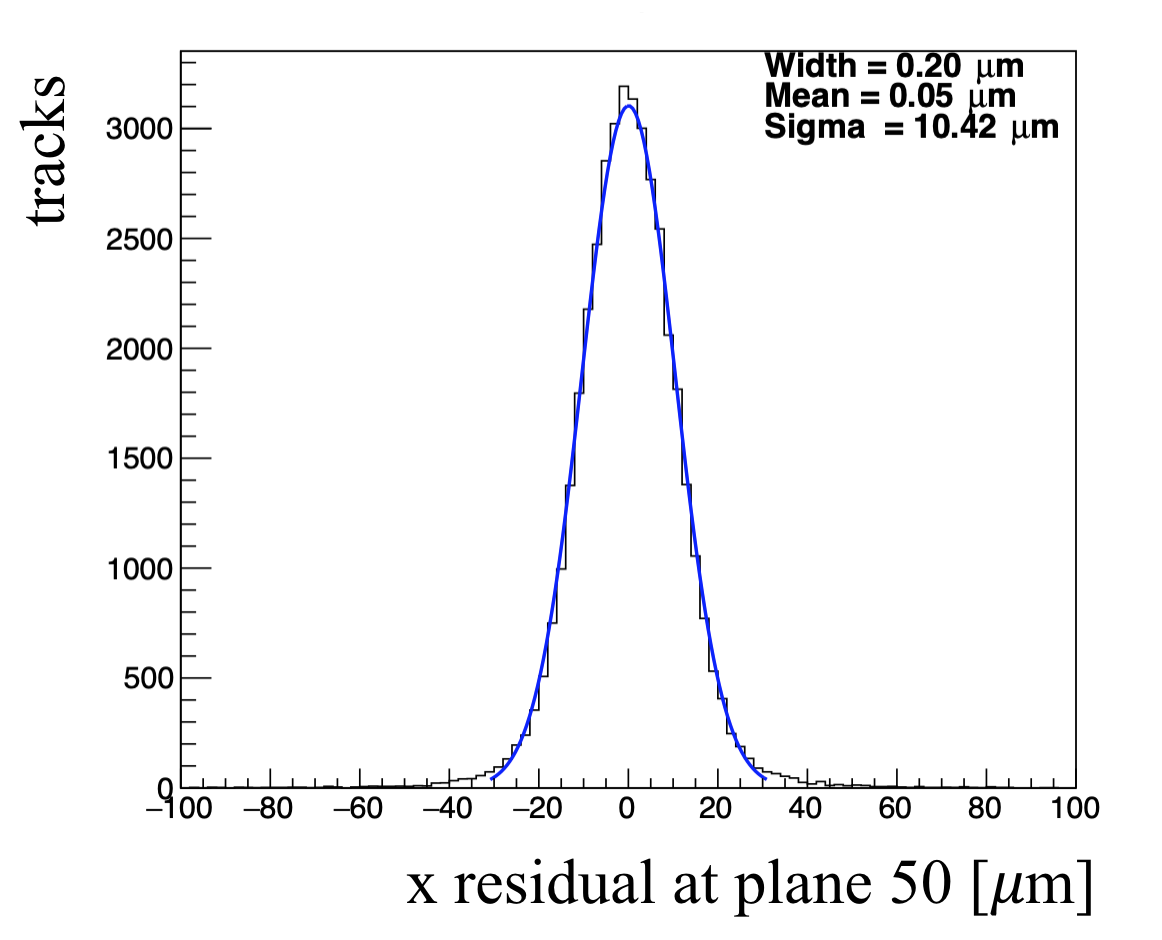}
}
\subfigure[]{
\label{fig:residual_RD53A_100_25:b} 
\includegraphics[width=0.45\textwidth]{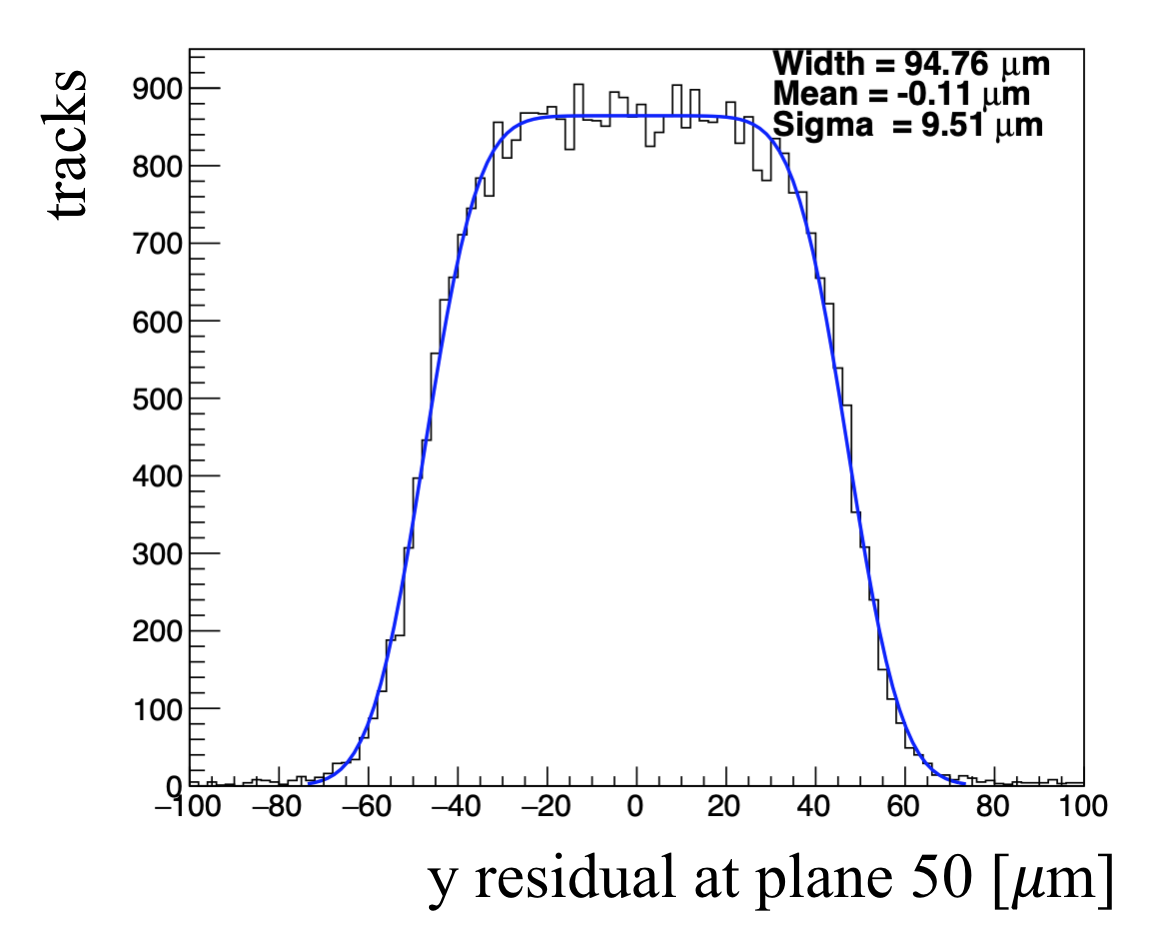}
}
\caption{(a) residual in X direction on $100\times25\ \mu$m$^2$ RD53A module. (b) residual in Y direction on $100\times25\ \mu$m$^2$ RD53A module.}
\label{fig:residual_RD53A_100_25} 
\end{figure}

As the hits on DUTs are excluded from the track reconstruction, the residuals on the DUTs are unbiased(Details in Appendix~\ref{apx:proof}):
\begin{equation}
\label{eq:unbiased_reso}
\sigma^{2}_{unbiased}=\sigma^{2}_{intrinsic}+\sigma^{2}_{track}.
\end{equation}
For a large pitch, the fit function should move to the convolution of box and Gaussian function. The box function has one parameter $d$ which is about equal to the pitch, and the Gaussian function has two free parameters, mean ($\mu$) and width ($\Sigma$). The unbiased resolution can be calculated by these fit parameters: 
\begin{equation}
\sigma^{2}_{unbiased}=d^2/12+\Sigma^{2}.
\end{equation}
The residuals on $100\times25\ \mu$m$^2$ RD53A module are shown on Figure~\ref{fig:residual_RD53A_100_25}.

\begin{figure}[!htb]
\centering
\includegraphics[width=\textwidth]{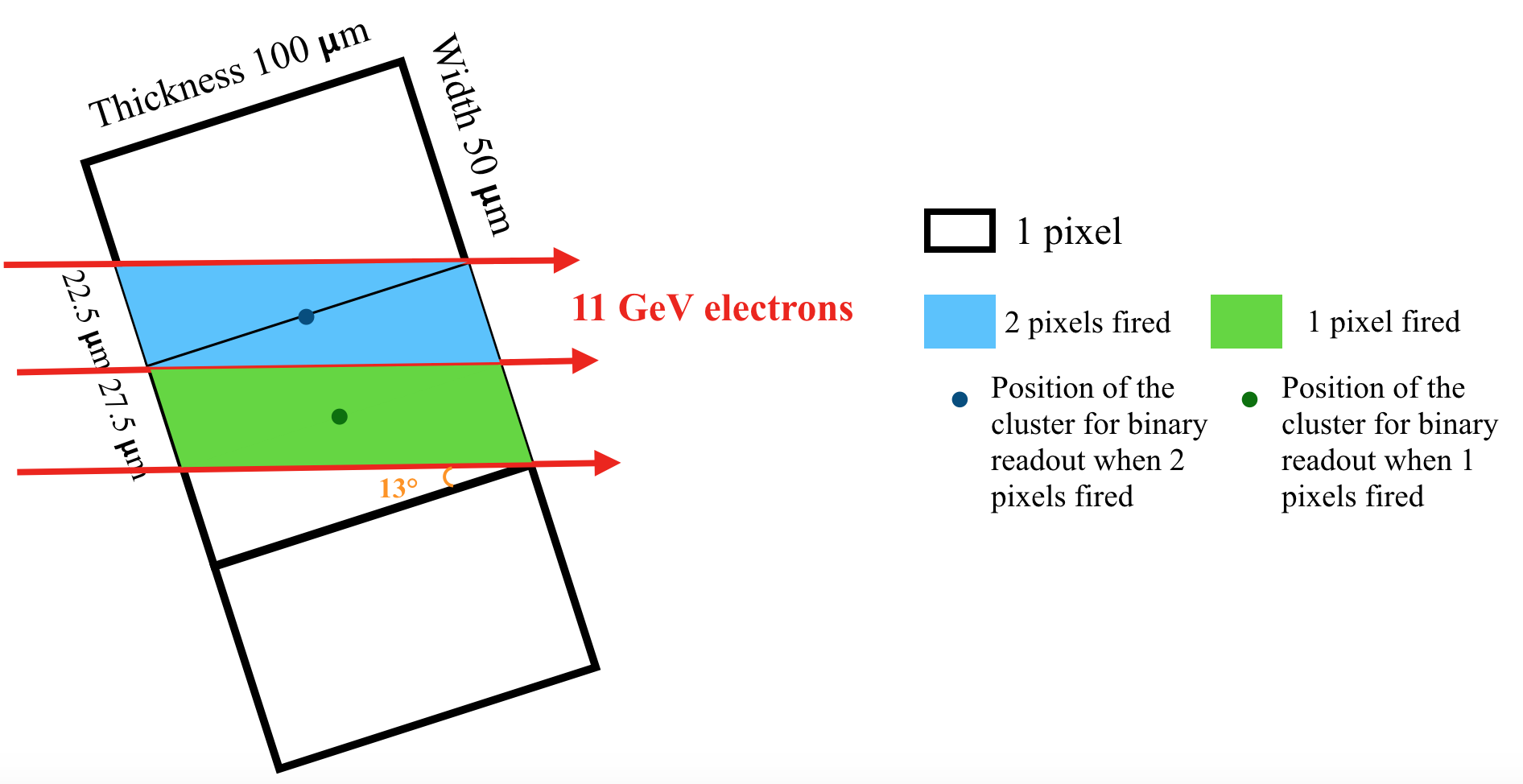}
\caption{Plot to demonstrate the effect from the tilt angle. The example pixel is 100 $\mu$m thick and 50 $\mu$m wide. The module is tilted by $13^\circ$.}
\label{fig:tilt_angle_effect}
\end{figure}

\begin{figure}[!htb]      
\centering
\subfigure[]{
\label{fig:residual_RD53A_50_50:a} 
\includegraphics[width=0.45\textwidth]{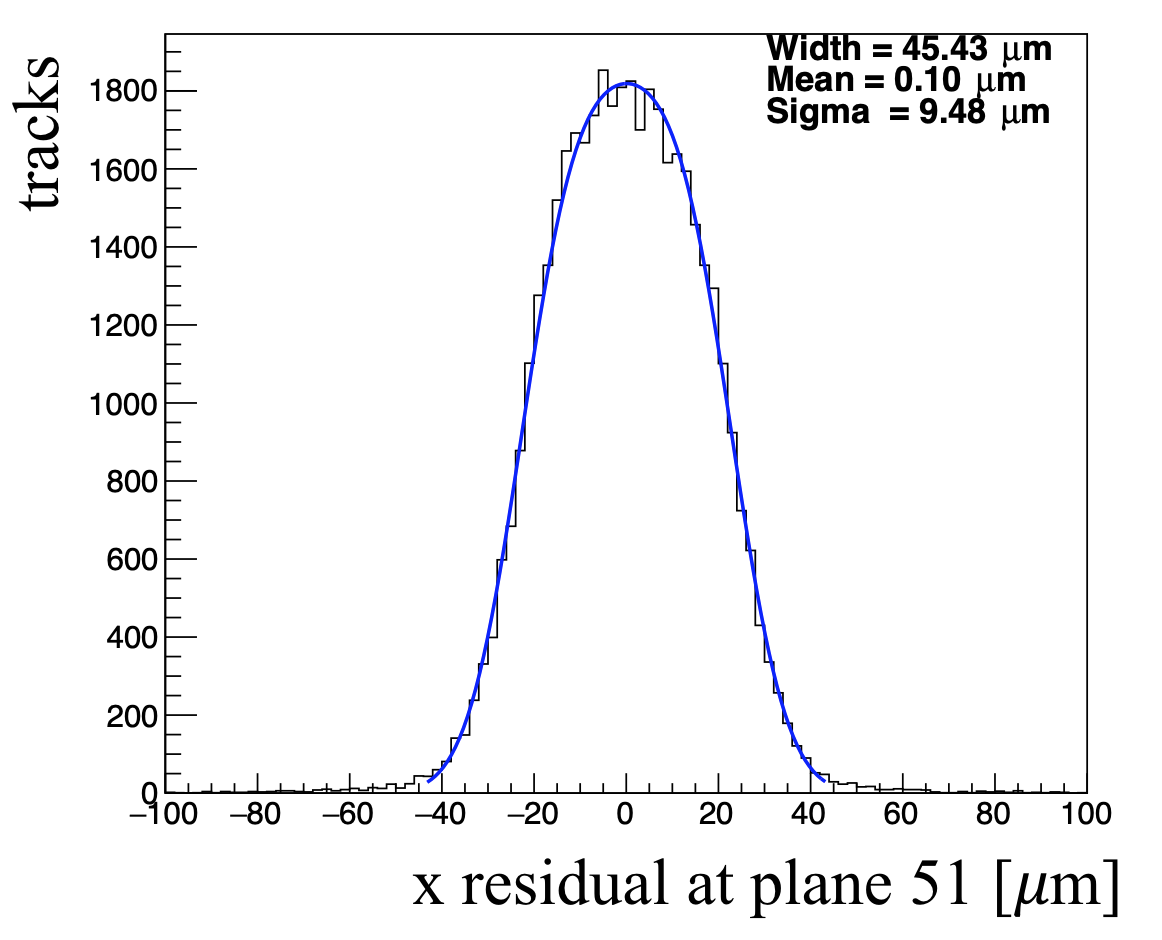}
}
\subfigure[]{
\label{fig:residual_RD53A_50_50:b} 
\includegraphics[width=0.45\textwidth]{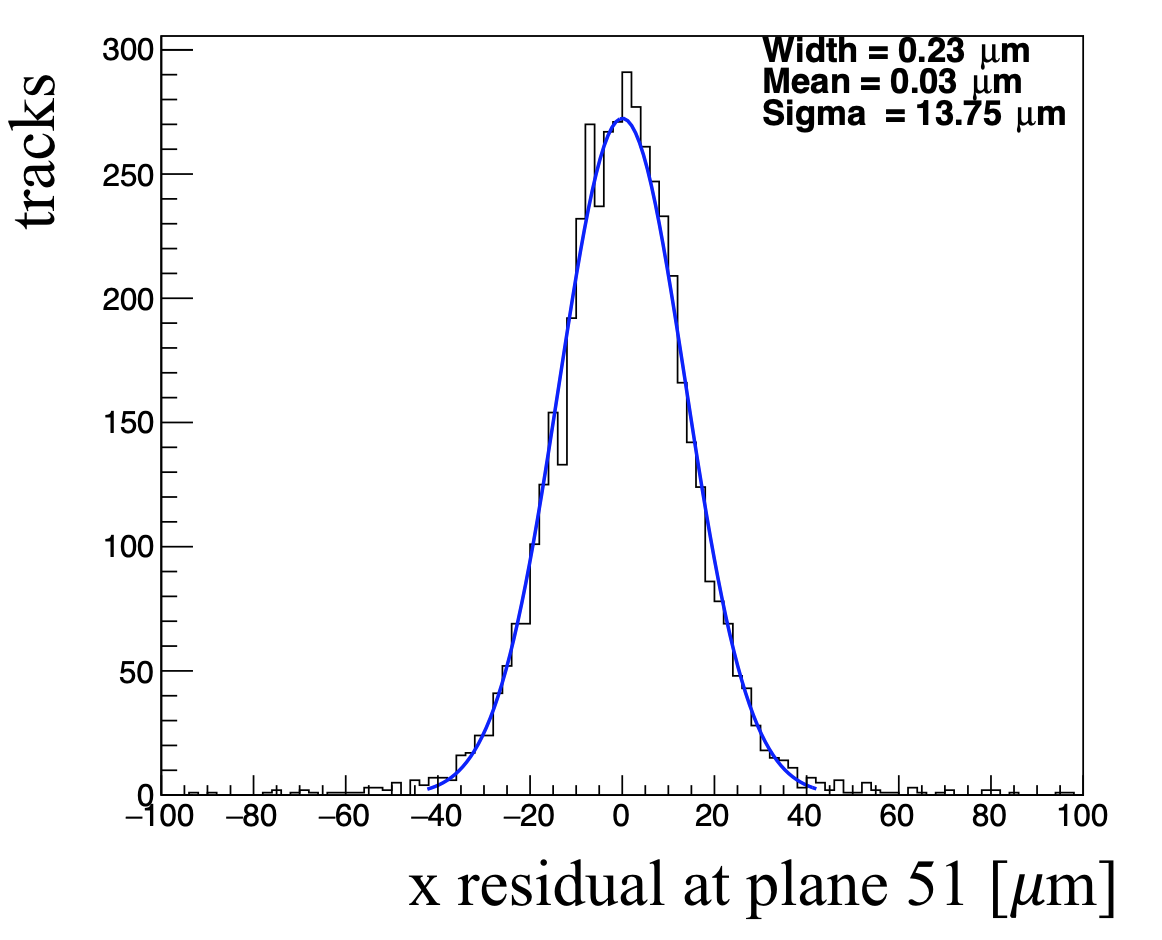}
}
\caption{(a) residual in X direction on $50\times50\ \mu$m$^2$ RD53A module. (b) residual in X direction on tilted $50\times50\ \mu$m$^2$ RD53A module.}
\label{fig:residual_RD53A_50_50} 
\end{figure}

If the DUTs are tilted by $13^\circ$, the equivalent pitch is reduced, which means a better position resolution, as depicted in Figure~\ref{fig:tilt_angle_effect}. When the incident particle fires two pixels, the exact position of the particle in blue area is not resolvable, which leads to 22.5 $\mu$m equivalent pitch. The equivalent pitch also shrinks to 27.5 $\mu$m for one pixel fired. Figure~\ref{fig:residual_RD53A_50_50} shows clearly that the position resolution for tilted DUT is better.

\begin{figure}[!htb]
\centering
\includegraphics[width=\textwidth]{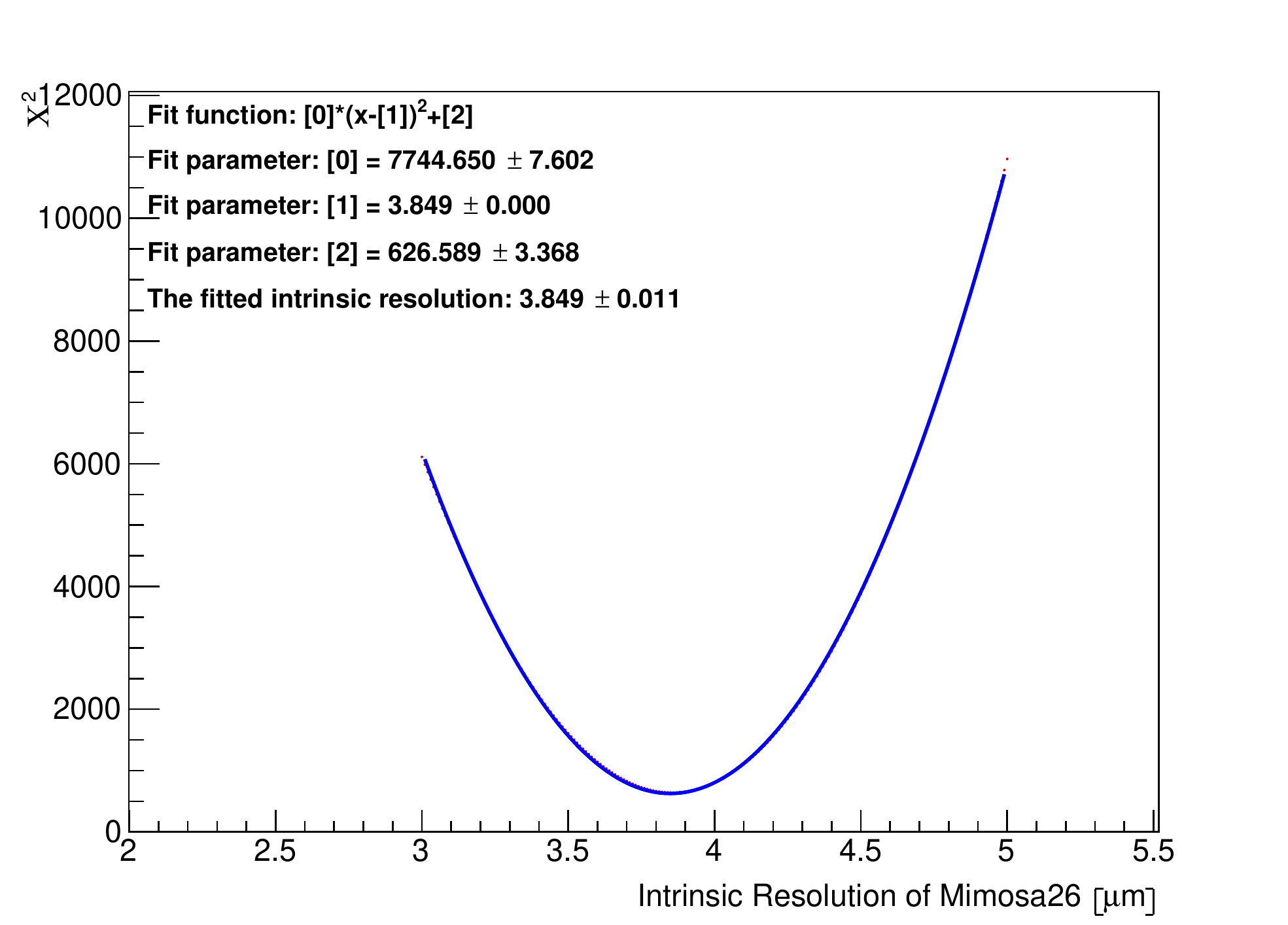}
\caption{$\chi^2$ distribution. Red dots are $\chi^2$ for each scanned intrinsic resolution of Mimosa26. Blue curve is the fitted parabola.}
\label{fig:chi2}
\end{figure}

\begin{figure}[!htb]
\centering
\includegraphics[width=\textwidth]{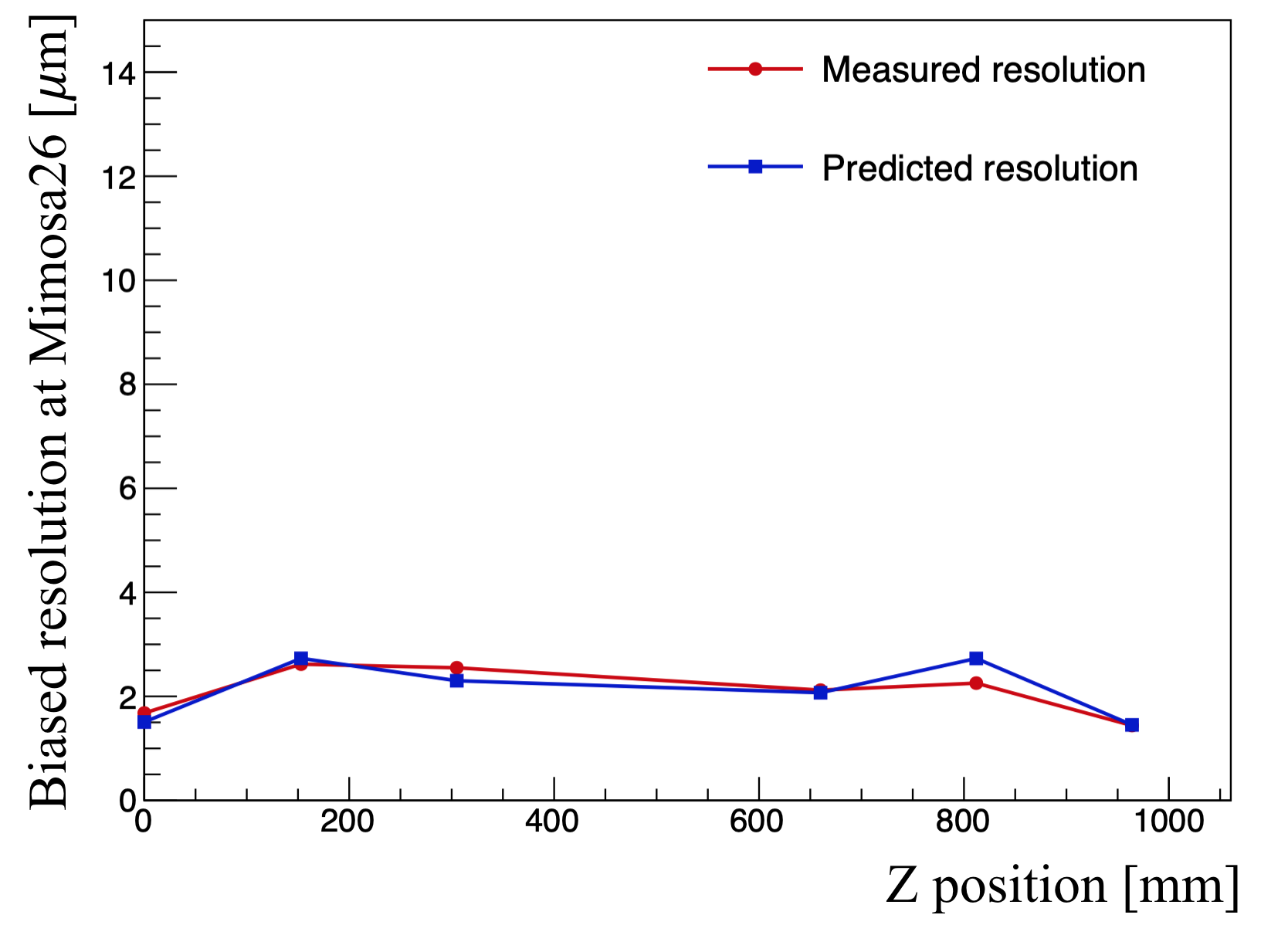}
\caption{Comparison between measured and predicted biased resolution on six Mimosa26 planes based on 3.85 $\mu$m intrinsic resolution.}
\label{fig:reso_comparison}
\end{figure}

In the testbeam measurement, only unbiased position resolutions are measured for DUTs, however, intrinsic resolutions are desired. Intrinsic resolution can be calculated by Equation (\ref{eq:unbiased_reso}) if the track resolution is known. Track resolutions can be simulated by a track resolution simulator~\cite{track-resolution-simulator}. In this package, the same geometry and materials as the ones used in EUTelescope should be provided. Moreover, the intrinsic resolution of Mimosa26 is also necessary. Although the intrinsic resolution of some EUDET-type telescopes have been already measured precisely~\cite{intrinsic_reso}, the intrinsic resolution of CALADIUM is unknown. Fortunately, The biased and unbiased resolution on Mimosa26 measured in SLAC testbeam can be used to determine the intrinsic resolution. To estimate the intrinsic resolution, the possible range (3 $\mu$m, 5 $\mu$m) is scanned. For each value with step size 0.01 $\mu$m, the following $\chi^2$ is calculated:
 \begin{equation}
\chi^2=\sum_{i}\frac{(\sigma_{biased, i}-\sqrt{\sigma^{2}_{intrinsic}-\sigma^{2}_{track,i}})^2}{V[\sigma_{biased, i}]},
\end{equation}
where $i$ denotes $i^{th}$ Mimosa26 plane. $\sigma_{intrinsic}$ is the scanned intrinsic resolution, and $\sigma_{track,i}$ is the track resolution on $i^{th}$ Mimosa26 plane simulated by the track resolution simulator with this intrinsic resolution. $\sigma_{biased, i}$ and $V[\sigma_{biased, i}]$ are the biased resolution and variance, respectively, extracted from fitting the biased residuals on Mimosa26. Note that the estimated resolution used in track fitting in EUTelescope are always 4.5 $\mu$m for Mimosa26 and $pitch/\sqrt{12}$ for DUTs. The $\chi^2$ distribution is well-fit by a parabola function, as shown in Figure~\ref{fig:chi2}. The value leading to the smallest $\chi^2$, 3.85 $\mu$m, is taken as the intrinsic resolution of all Mimosa26. Figure~\ref{fig:reso_comparison} shows the good agreement between measured and predicted biased resolution on six Mimosa26 planes based on 3.85 $\mu$m intrinsic resolution.

Once the intrinsic resolution of Mimosa26 is determined, the track resolution on each plane can be obtained from the track resolution simulator. Then, the intrinsic resolution of DUTs can be derived from Equation (\ref{eq:unbiased_reso}). Table~\ref{tab:result} summarizes the intrinsic position resolution of two RD53A modules for tilted and non-tilted cases, which are also compared with $pitch/\sqrt{12}$. The following systematic uncertainties are considered in this measurement:
\begin{itemize}
	\item The material on DUT is hard to estimate precisely, and it is critical for multiple scattering. The configuration of material in EUtelescope varies by 10\% as the systematic uncertainty.
	\item The beam energy is also varied by 10\%.
	\item The position of DUTs in Z direction is measured with poor accuracy, thus, 10 mm uncertainty is given.
	\item The intrinsic resolution of Mimosa26 described above is extracted from biased residuals. Actually, the unbiased residual of each Mimosa26 plane are also available when the corresponding plane is excluded in track fitting. The intrinsic resolution of Mimosa26 is measured as 3.35 $\mu$m using unbiased residuals, which results in some systematic uncertainty for track resolution on DUTs. 
\end{itemize}

\renewcommand\arraystretch{2}
\begin{table}[!htb]
\centering
\resizebox{\textwidth}{!}{
\begin{tabular}{c|cccc} \hline 
& \makecell{$50\times50\ \mu$m$^2$ RD53A\\ non-tilted side(50 $\mu$m)} & \makecell{$50\times50\ \mu$m$^2$ RD53A\\ tilted side(50 $\mu$m)} & \makecell{$100\times25\ \mu$m$^2$ RD53A\\ non-tilted side(100 $\mu$m)} & \makecell{$100\times25\ \mu$m$^2$ RD53A\\ tilted side(25 $\mu$m)} \\ \hline
$pitch/\sqrt{12}$				&	14.4			&	14.4			&	28.8			&	7.2		   	\\
 Non-tilted DUTs				&	$14.51\pm1.05$	&	$14.58\pm1.04$	&	$28.16\pm0.67$	&	$7.92\pm1.73$	\\ 	
$13^\circ$ tilted DUTs			&	$14.04\pm1.07$	&	$10.86\pm1.09$	&	$28.54\pm0.75$	&	$6.81\pm1.82$	\\ 	
 $\frac{13^\circ\ tilted}{Non-tilted}$	&	$0.97\pm0.10$	&	$0.74\pm0.09$	&	$1.01\pm0.04$	&	$0.86\pm0.30$	\\  \hline
\end{tabular}}
\caption{Intrinsic position resolution of tilted and non-tilted RD53A modules.}
\label{tab:result}
\end{table}

\section{Conclusion}

The intrinsic position resolution of non-tilted and tilted RD53A modules with $50\times50$ and $25\times100\ \mu$m$^2$ pitch are measured using a 11 GeV electron beam at SLAC. The intrinsic resolution of non-tilted RD53A modules are both comparable with $pitch/\sqrt{12}$, which reduces by 26\% and 14\% respectively for $13^\circ$ tilted $50\times50$ and $25\times100\ \mu$m$^2$ RD53A. This result is useful for deciding on the geometry of the pixel layers in phase \uppercase\expandafter{\romannumeral2} upgrade, which are crucial for flavor tagging and many other tasks.

\section*{Acknowledgements}
This work is supported by US ATLAS and the U.S.~Department of Energy, Office of Science under contract DE-AC02-05CH11231.  We would like to thank all the shifters who participated in the data collection (Sanha Cheong, Aleksandra Dimitrievska, Su Dong, Timon Heim, Charilou Labitan, Ke Li, Maurice Garcia-Sciveres, Simone Mazza, Patrick McCormak, Elisabetta Pianori, Cesar Renteria, Mark Standke, Adrien Stejer, Zijun Xu) as well as the SLAC ESTB and accelerator staff (especially Carsten Hast and Toni Smith). We also thank Hendrik Jansen and Simon Spannagel for helpful discussions on the resolution simulation as well as comments on the manuscript. Finally, we are grateful to extensive help on EUDAQ and EUTelescope from the ATLAS ITk community and beyond.

\clearpage

\appendix
\renewcommand\thefigure{\Alph{section}\arabic{figure}}
\renewcommand\thetable{\Alph{section}\arabic{table}}
\renewcommand\theequation{\Alph{section}\arabic{equation}}

\section{The relationship between (un)biased residuals, intrinsic resolution and track resolution}
\label{apx:proof}
\setcounter{figure}{0}
\setcounter{table}{0}
\setcounter{equation}{0}

\noindent  Suppose $N$ pixel planes where on each plane we measure a position $\vec{y}$ transverse to the beam. Plane $i$ is located at location $z_i$.   Some subset of these pixel planes $D$ are special and called devices under test (DUT) and the other ones are called telescope planes $T$.  A \textit{track} is a function $f(z|\theta)$ constructed to model the trajectory of a charged particle traversing the pixel planes, where $\theta$ are parameters that can be fit to the data.  In particular,

\begin{align}
\theta_\text{unbiased}&=\text{argmin}_{\theta'} \sum_{i\in T} \mathcal{L}(f(z_i|\theta'),\vec{y}_i,\Sigma_i)\\
\theta_\text{biased}&=\text{argmin}_{\theta'} \sum_{i\in T\cup D} \mathcal{L}(f(z_i|\theta'),\vec{y}_i,\Sigma_i),
\end{align}

\noindent where $\mathcal{L}$ is some loss function that goes to zero as $f(z_i|\theta')\rightarrow \vec{y}_i$ and $\Sigma_i$ is the covariance matrix corresponding to the uncertainty on measurement $i$.  In particular, 

\begin{align}
\Sigma_i=\langle (\vec{y}_i-\vec{y}_i^\text{true})(\vec{y}_i-\vec{y}_i^\text{true})^T\rangle,
\end{align}

\noindent where $\vec{y}_i^\text{true}$ is where the actual particle went through plane $i$.   The \textit{intrinsic} resolution of plane $i$ is defined to be $\Sigma_i$.  We need two more quantities.  The (un)biased \textit{track resolution} at a location $z$ is defined to be

\begin{align}
\Sigma_\text{(un)biased track}(z) = \langle (f(z|\theta_\text{(un)biased})-\vec{y}^\text{true}(z))(f(z|\theta_\text{(un)biased})-\vec{y}^\text{true}(z))^T\rangle,
\end{align}

\noindent where $\vec{y}^\text{true}(z)$ is the actual position of the particle at location $z$ (in particular, $\vec{y}_i^\text{true}=\vec{y}^\text{true}(z_i)$).  Finally, the (un)biased \text{residual} is defined to be $r_i=f(z_i|\theta)-\vec{y}_i$.  The residual resolution is:

\begin{align}
\Sigma_\text{(un)biased residual,$i$} = \langle (r_i-\langle r_i\rangle) (r_i-\langle r_i\rangle)^T\rangle
\end{align}

\vspace{2mm}

\begin{figure}[!htb]
\centering
\includegraphics[width=0.5\textwidth]{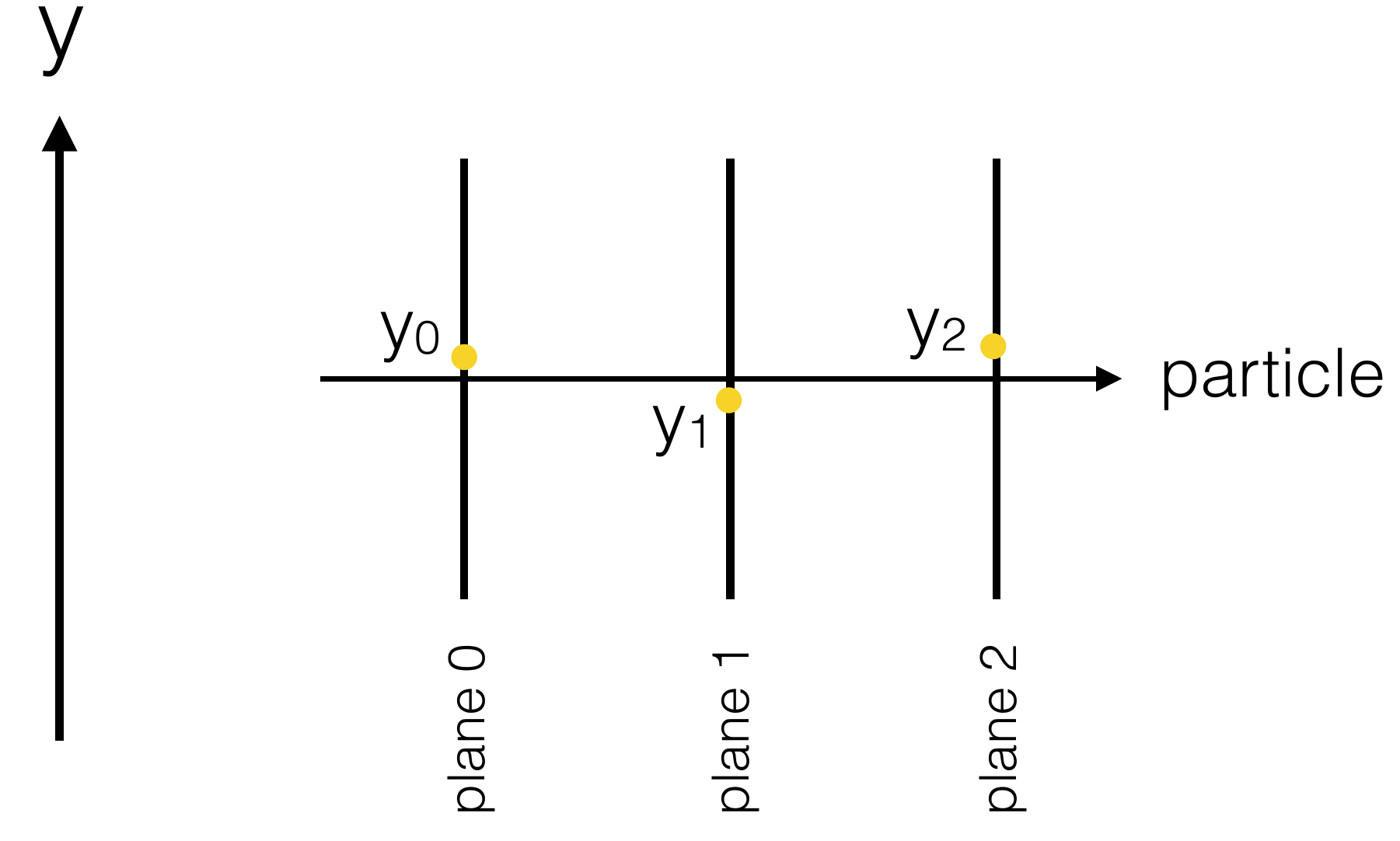}
\caption{The setup for the simple calculation.}
\label{fig:blah}
\end{figure}

\noindent To make our lives easier, let's consider the 1D case, as shown in Fig.~\ref{fig:blah} so that $\vec{y}_i$ is just a number $y_i$ and $\Sigma$ is also just a number $\sigma^2$. The claim is that these two formulae are true:

\begin{align}
\sigma^2_\text{unbiased residual,$i$}&=\sigma^2_\text{intrinsic,$i$}+\sigma^2_\text{unbiased track}\\
\sigma^2_\text{biased residual,$i$}&=\sigma^2_\text{intrinsic,$i$}-\sigma^2_\text{biased track}
\end{align}

To further simplify things, let's take $N=3$, $D=\{1\}$, $T=\{0,1\}$, $\mathcal{L}(f(z_i|\theta),y_i,\sigma_i)=(f(z_i|\theta)-y_i)^2/\sigma_i^2$, $\sigma_0=\sigma_2$, $\langle y_i\rangle=y_i^\text{true}$, and $\partial_z f=0$.  In other words, we have three planes, two that are the telescope, and we are going to fit a straight line $f=\theta$, a constant (no slope - perpendicular incidence and no multiple scattering).

\clearpage

\noindent All of the above quantities can be computed in our simple example.  First, let's compute some tracks:

\begin{align}
\theta_\text{unbiased}&=\text{argmin}_{\theta'} \left(\frac{(y_0-\theta')^2}{\sigma_0^2}+\frac{(y_2-\theta')^2}{\sigma_2^2}\right) \\
&=\frac{y_0+y_2}{2}
\end{align}

\begin{align}
\theta_\text{biased}&=\text{argmin}_{\theta'} \left(\frac{(y_0-\theta')^2}{\sigma_0^2}+\frac{(y_1-\theta')^2}{\sigma_1^2}+\frac{(y_2-\theta')^2}{\sigma_2^2}\right) \\
&=\frac{\frac{y_0}{\sigma_0^2}+\frac{y_1}{\sigma_1^2}+\frac{y_2}{\sigma_2^2}}{\frac{1}{\sigma_0^2}+\frac{1}{\sigma_1^2}+\frac{1}{\sigma_2^2}}
\end{align}

\noindent Next, compute the track resolution at $z=z_1$:

\begin{align}
\sigma^2_\text{unbiased track}&=\langle(\theta_\text{unbiased}-y_1^\text{true})^2\rangle \\
&=\left\langle\left(\frac{y_0+y_2}{2}-y_1^\text{true}\right)^2\right\rangle\\
&=\left\langle\left(\frac{y_0-y_0^\text{true}+y_2-y_2^\text{true}}{2}\right)^2\right\rangle\\
&=\frac{1}{4}(\sigma_0^2+\sigma_2^2)
\end{align}

\begin{align}
\sigma^2_\text{biased track}&=\langle(\theta_\text{biased}-y_1^\text{true})^2\rangle \\
&=\frac{1}{\frac{1}{\sigma_0^2}+\frac{1}{\sigma_1^2}+\frac{1}{\sigma_2^2}}
\end{align}

\clearpage

\noindent Next, let's compute the residuals:

\begin{align}
r_\text{unbiased,$i$}&=y_1-\frac{y_0+y_2}{2}\\
r_\text{biased,$i$}&=y_1-\frac{\frac{y_0}{\sigma_0^2}+\frac{y_1}{\sigma_1^2}+\frac{y_2}{\sigma_2^2}}{\frac{1}{\sigma_0^2}+\frac{1}{\sigma_1^2}+\frac{1}{\sigma_2^2}}
\end{align}

\noindent Note that with our assumptions, the average residual is zero for both cases. The biased and unbiased residual resolution just follow the above results:

\begin{align}
\sigma^2_\text{unbiased residual,$i$}&=\langle (r_\text{unbiased,$i$}-\langle r_\text{unbiased,$i$}\rangle)^2\rangle\\
&=\left\langle \left(r_\text{unbiased,$i$}-\left\langle y_1-\frac{y_0+y_2}{2}\right\rangle\right)^2\right\rangle\\
&=\left\langle \left(r_\text{unbiased,$i$}-y_1^\text{true}+\frac{y_0^\text{true}+y_2^\text{true}}{2}\right)^2\right\rangle\\
&=\left\langle \left(y_1-y_1^\text{true}+\frac{(y_0^\text{true}-y_0)+(y_2^\text{true}-y_2)}{2}\right)^2\right\rangle\\
&=\sigma_1^2+\frac{1}{4}(\sigma_0^2+\sigma_2^2)\\
&=\sigma^2_\text{intrinsic,$i$}+\sigma^2_\text{unbiased track}
\end{align}

\begin{align}
\sigma^2_\text{biased residual,$i$}&=\langle (r_\text{biased,$i$}-\langle r_\text{biased,$i$}\rangle)^2\rangle\\
&=\left\langle \left(y_1-\frac{\frac{y_0}{\sigma_0^2}+\frac{y_1}{\sigma_1^2}+\frac{y_2}{\sigma_2^2}}{\frac{1}{\sigma_0^2}+\frac{1}{\sigma_1^2}+\frac{1}{\sigma_2^2}}\right)^2 \right\rangle\\
&=\sigma_1^2+\frac{\frac{1}{\sigma_1^2}}{\left(\frac{1}{\sigma_0^2}+\frac{1}{\sigma_1^2}+\frac{1}{\sigma_2^2}\right)^2}-\frac{2}{\frac{1}{\sigma_0^2}+\frac{1}{\sigma_1^2}+\frac{1}{\sigma_2^2}}+\frac{\frac{1}{\sigma_0^2}+\frac{1}{\sigma_2^2}}{\left(\frac{1}{\sigma_0^2}+\frac{1}{\sigma_1^2}+\frac{1}{\sigma_2^2}\right)^2}\\
&=\sigma^2_\text{intrinsic,$i$}-\sigma^2_\text{biased track}
\end{align}
 
\end{document}